\documentclass[twocolumn,showpacs,preprintnumbers,amsmath,amssymb]{revtex4}

\usepackage{graphicx}% Include figure files
\usepackage{dcolumn}% Align table columns on decimal point
\usepackage{bm}% bold math
\usepackage{color}
\usepackage[draft,english]{fixme}

\begin{document}

\title{Double Neutrino Production and Detection in Neutrino Detectors}
\author{Don van der Drift} \affiliation{Lawrence Berkeley National Laboratory, Berkeley CA, 94720, USA and the University of California, Berkeley, CA, 94720, USA}
\author{Spencer R. Klein} \affiliation{Lawrence
Berkeley National Laboratory, Berkeley CA 94720 USA,  and the University of California, Berkeley, CA, 94720, USA}

\begin{abstract}

Large, high-energy ($E>100$ GeV) cosmic neutrino telescopes are now quite mature. IceCube, for example, observes about 50,000 well-reconstructed single atmospheric neutrino events/year, with energies above 100 GeV.  Although the neutrino detection probability is small, current detectors are large enough so that it is possible to detect two neutrinos from the same cosmic-ray interaction.  In this paper, we calculate the expected rate of double-neutrino interactions from a single cosmic-ray air shower. The rate is small, about 0.07 events/year for a 1 km$^3$ detector like IceCube, with only a small dependence on the assumed cosmic-ray composition and hadronic interaction model.  For a larger detector, like the proposed KM3Net, the rate is about 0.8 events/year, a rate that should be easily observable.  These double neutrino interactions are the major irreducible background to searches of pairs of particles produced in supersymmetric neutrino or cosmic-ray air-shower interactions.   Other standard-model backgrounds are considered, and found to be small. 

\end{abstract}

\maketitle

\section{Introduction}

High-energy astrophysical neutrinos are being actively studied, with a view to using them to find the acceleration sites for high-energy cosmic rays \cite{Halzen:2008zz,Gaisser:1994yf}. Three large detectors, the 1 km$^3$ IceCube \cite{IceCube}, the $10^{7}$ m$^3$ Baikal detector \cite{Dzhilkibaev:2009ja}, and the $1.5\times 10^7$ m$^3$ ANTARES detector \cite{Collaboration:2011nsa} are taking data, and the 5-6 km$^3$ KM3NeT detector \cite{Sapienza:2011zzb,Coniglione}  has been proposed. These detectors are optimized for neutrinos with energies in the TeV to PeV range, where the atmospheric neutrino flux is substantial; the completed IceCube detector, for example, observes about 50,000 well-reconstructed single atmospheric neutrino events/year \cite{Sullivan:2012hf}.  Most of these events are from $\nu_\mu$ (or $\overline\nu_\mu$; we do not distinguish between neutrinos and antineutrinos here), which interact and produce muons which travel upward through the detector. 

Although these detectors are focused on the detection of single neutrino interactions, they also look for more complex topologies. One signature of great interest consists of two parallel tracks going upward through the detector.   This signature could be a sign of some type of `new physics,' such as supersymmetry (SUSY) or Kaluza-Klein models.  In supersymmetry, parallel tracks can be created when a neutrino (or cosmic-ray) interacts in the Earth below a detector, producing a pair of SUSY particles \cite{Albuquerque:2006am,Helbing:2011wf}.  These supersymmetric particles decay, eventually producing a pair of next-to-lightest SUSY particles.  If SUSY has a high mass scale, then these particles have a relatively long lifetime, of order $\mu$s.  They live long enough to travel long distances ($\approx 1,000$ km) through the Earth.  During this trip, they will slowly separate, and will appear in a neutrino detector as a pair of upward-going parallel tracks, with a typical separation of order a few hundred meters \cite{Albuquerque:2009vk}.   Since these particles are typically quite heavy, they lose energy like nearly minimum ionizing particles.  Kaluza-Klein particles are produced via a different mechanism, but have similar observational consequences \cite{Albuquerque:2008zs}.  

Previous studies have considered the standard-model backgrounds to these processes; the major background is from charm production, where both of the charmed particles decay semileptonically \cite{Albuquerque:2006am}. This produces a pair of muons. These muons have a rather short range (even a 1 PeV muon has a range less than 10 km in rock), and a typical transverse momentum of a few GeV/c, so are unlikely to have separated significantly by the time they range out.

Two muons from a pair of neutrino interactions, from the same cosmic-ray air shower, are the only background that is likely to mimic the signatures described above. If the neutrinos are produced in the same cosmic-ray air shower, then they will be nearly parallel, but with a large enough opening angle to separate by a few hundred meters as they pass through the Earth. If the neutrinos have an energy below a few TeV, they may appear to be quasi-minimum-ionizing, and could mimic a pair of supersymmetric or Kaluza-Klein particles.
  
In this paper, we calculate the rate of double-neutrino events expected to be observable in IceCube and KM3NeT, and discuss the expected characteristics of the events \cite{thesis}.  We do not differentiate between $\nu$ and $\overline\nu$ here. 

\section{Air Showers, Production Model and neutrino interactions}

The calculation was done in two parts.  In the first part, cosmic-ray air showers are generated, and the neutrino data retained for analysis in the second part.  All the neutrinos in each event are paired with all of the other neutrinos in that event, and the separation distance computed. The neutrino-pair flux is weighted with the probability of detection.  The detection probability has  three components: the energy-dependent probabilities of the two neutrinos interacting with the resulting muon being observed (assumed independent for each neutrino), and the probability (based on their separation) of both muons passing within the detector active volume. The lateral separation distance $D$ is a critical parameter. It depends on the distance between the shower and the detector, and on the opening angle between the two neutrinos.   The opening angle depends on the neutrino energy and transverse momentum, $p_T$ (relative to the shower core).  This analysis is sensitive to the particles with the smallest $p_T$, unlike studies of high $p_T$ muons \cite{IceCube:2012ij}.   $\nu_\mu$ from pion and kaon decay dominate the atmospheric spectrum \cite{Aartsen:2012uu}, and so should constitute most of the signal; we focus on them.  We only consider neutrinos with energies above 100 GeV; because of their small interaction cross-section and large angular spread, lower energy neutrinos do not contribute significantly.   We assume that the neutrinos are produced by the decays of different pions.  

Cosmic-ray air showers were generated using CORSIKA version 6.980 \cite{CORSIKA}  to model the cosmic-ray air showers. Two different hadronic interaction models, QGSJET v01c  \cite{Kalmykov:1997te} and DPMJET v2.55 \cite{Ranft:1999qe} were used. The cosmic-ray spectrum was approximated by the H\"orandel spectrum \cite{Hoerandel:2002yg}, with the low-energy end of the cosmic-ray spectrum following an $E^{-2.7}$ slope, and the high-energy end following $E^{-3}$:
\begin{equation}
	\Phi(E_p)=
	\begin{cases}
		1.8\cdot 10^4 E_p^{-2.7} & Ep < 10^6 \text{ GeV}\\
		1.1\cdot 10^6 E_p^{-3.0} & Ep > 10^6 \text{ GeV}
	\end{cases}\label{eq:Flux},
\end{equation}
with $E_p$ the energy of the primary particle in units of GeV and the flux, $\Phi(E_p)$ in 1/(s sr m\textsuperscript{2} GeV).  

CORSIKA includes the bending effect of the Earths magnetic field, and multiple scattering in the Earths atmosphere.  Multiple scattering is negligible, but the magnetic bending of the $\pi^\pm$ and $K^\pm$ that decay into $\nu_\mu$  affects the calculation.   For a $5*10^{-4}$ T field (typical for the sky above Antarctica) perpendicular to the pion direction of motion, a pion is bent by an angle $\theta_B = qB c\tau_\pi/m_\pi\approx {\rm 117 keV}/m_\pi \approx 8.4\times10^{-4}$, where $m_\pi$ and $\tau_\pi$ are the $\pi^\pm$ mass and lifetime respectively.   The actual angle between the pion direction and the magnetic field is usually less than $90^0$, so the magnetic bending will be smaller, typically by of order $1/\sqrt{2}\approx 0.7$.

In comparison, the bending due to the pion/kaon transverse momentum, $p_T$  with respect to the cosmic-ray direction is $\theta_P = p_T/E_\pi$.  For a typical scale $\Lambda_{\rm QCD} = 300 $ MeV, the magnetic bending is larger than the $p_T$ induced bending for pion energies above 500 GeV.  For kaons at the same energy, the bending is a factor of 8 smaller.   The typical neutrino energy is around 1 TeV, so magnetic bending is important.   Because of the magnetic bending, the separation requirement preferentially selects like-sign pion pairs, so that the event sample will prefer neutrino-neutrino and antineutrino-antineutrino pairs, rather than mixed pairs.

The probability for two neutrinos to pass through a finite-sized detector depends on their transverse momentum ($p_T$) with respect to each other.  The separation $d$ depends on the relative $p_T$,  the particle energies, and the distance travelled. 
Since these neutrinos travel hundreds to thousands of kilometers before interacting, the two-neutrino rate is sensitive to the production of particles with very low $p_T$.   Both QGSJET and DPMJET generate low-$p_T$ particles using phenomenological, Pomeron-based models.  Both of them reproduce accelerator data quite well, and so have similar $p_T$ spectra in this region.  These neutrinos mostly come from the decay of pions and kaons, and the muons that are produced from the pion/kaon decays.  In the low $p_T$ region, both models predict thermal spectra that are in agreement with experimental data obtained in accelerator experiments.    Collider experiments are not sensitive at very low $p_T$, so there is some uncertainty here, but we can use data on high-energy cosmic-ray muon separations to check the models.
MACRO \cite{Ambrosio:1999qu} and IceCube \cite{IceCube:2012ij} have studied muon separation spectra at small and large separations respectively.  The observed separation spectra and overall rates are in reasonable agreement with Monte Carlo expectations, although the zenith angle distributions do not agree well.  

\begin{figure}[htbp]
\begin{center}
\includegraphics[width=0.5\textwidth]{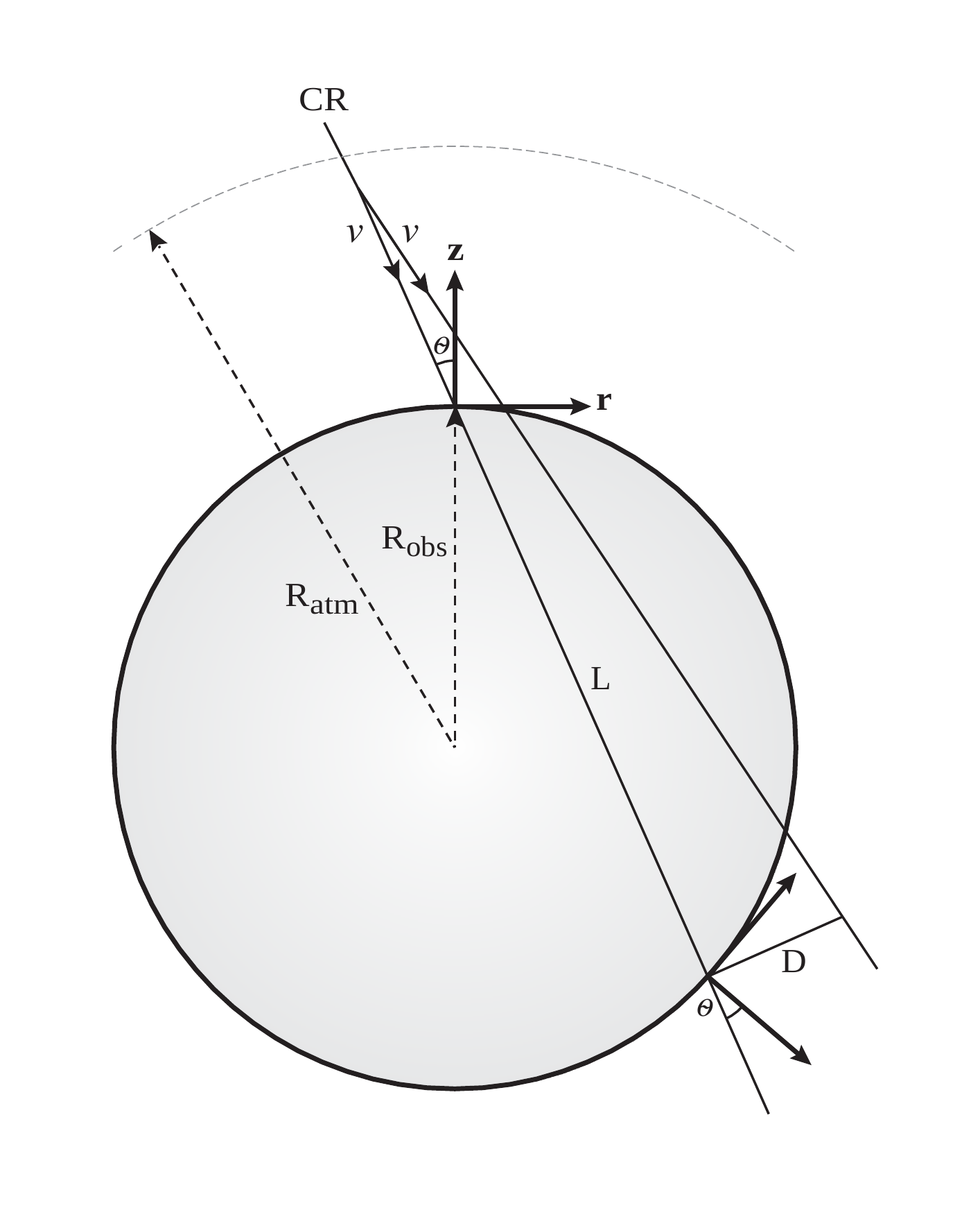}
\caption{The geometry used in the calculation for an incident cosmic ray at zenith angle $\theta$. The parameters used are: $R_{obs}$ for the radius from the center of the Earth to the observation height, at which the particles are saved, $R_{atm}$ for the radius form the center of the Earth to the top of the atmosphere, $L$ the path length through the Earth and $D$ the final separation after propagation through the Earth.}
\label{fig:geometry}
\end{center}
\end{figure}

CORSIKA generates downward-going showers; for this analysis, we used the transformation shown in Fig. \ref{fig:geometry} to convert the downward-going neutrinos into upward-going.  The transformation maps the zenith angle, $\theta_Z$ into $-\theta_Z$.   The two neutrinos are propagated through the Earth, separating as they go.  In the relevant energy range (100 GeV to 10 TeV), neither neutrino oscillation nor absorption in the Earth is significant.   

One weakness of this approach is that it uses both the magnetic field and ground elevation at the South Pole for all showers.   Most of the relevant showers occur within 1000 km of the detector, and estimates of the inaccuracy due to the simplification need only consider field variations over this distance scale.  For IceCube, we consider the region south of latitude $-75^0$.     Although the magnetic field strength does not vary significantly there, its direction does.   The dip angle (angle between the magnetic field lines and vertical) ranges from $-65^0$ to $-78^0$ there.  For a given longitude, the maximum change is $6^0$ as the latitude varies from  $-75^0$ to $-90^0$ \cite{Bfield} .  The declinations vary more, up to $30^0$.   The field variations can alter the magnetic bending by up to 50\%, but, after averaging over all possible angles of incidence, the net effect will be much smaller; the overall change in rate should be less than 25\%.  

The opening angle is bounded by the geometry, the observed perpendicular separation when the neutrinos are saved by the simulation and the maximum separation allowed inside the detector. The minimum is set by the observed separation, assuming that the neutrino separation started within the atmosphere.  The maximum is determined by the path length through the Earth and the subsequent increase in separation.

The neutrino detection probability depends on both the neutrino interaction probability and the probability of observing the produced muon.  We use a simple model which includes both factors \cite{Gaisser:1994yf}
\begin{equation}
	{\rm P}({\rm detection|E_{\nu i}}) = 
	\begin{cases}
    		1.3\:10^{-6}\:E_\nu^{2.2} 	& \text{if $E_\nu \leq 1$ TeV}\\
    		1.3\:10^{-6}\:E_\nu^{0.8} 	& \text{if $E_\nu > 1$ TeV}\\
		0 					& \text{if $D > D_{max}$}\\
	\end{cases}\label{eq:ProbDet},
\end{equation}
with $E_\nu$ the neutrino energy in units of TeV.  

These probabilities are based on the model of a detector as a thin plate, sensitive to muons (from $\nu_\mu$ and $\overline\nu_\mu$ interactions).  Neutrinos are detected if the muons that they produce have a sufficient range to reach the plate.

This calculation neglects the minimum separation for two muons to be detectable as separate particles, $D_{\rm min}$. 
Reconstruction of two parallel tracks is more challenging than for single tracks, with additional degrees of freedom \cite{Ribordy:2006qd}.   The IceCube collaboration found that downward-going isolated muons were separable from muon bundles at separations larger than 135 m \cite{IceCube:2012ij}.   For two single muons, the minimum observable separation could be somewhat lower, especially for near-horizontal muon pairs.  The rate correction scales as $(D_{\rm min}/D_{\rm max})^2$.  For $D_{\rm min}=135$ m, and $D_{\rm max} = 1$ km, as in IceCube, this is a small correction; for KM3NeT, it would be even smaller. 

Both Icecube and KM3NeT are 3-dimensional, so that neutrinos that interact anywhere in the detector volume may be observed, in addition to neutrinos that interact outside the detector, but whose muons reach the detector volume. Both detectors contain holes, regions where a low-energy (minimum ionizing) muon may pass through undetected.   As the neutrino energy rises, the muon range and energy loss both rise, and both effects become less important.   Over the relevant neutrino energies, these effects are both less than a factor of two.  Fortunately, they work in different directions, and we will assume that they will cancel out.  A more accurate calculation would require a detailed dedicated detector Monte Carlo to account for the correlated detection probability, event reconstruction software, and a well defined set of event selection criteria. 

\section{Results}

\begin{figure}[htb]
\begin{center}
\includegraphics[width=0.5\textwidth]{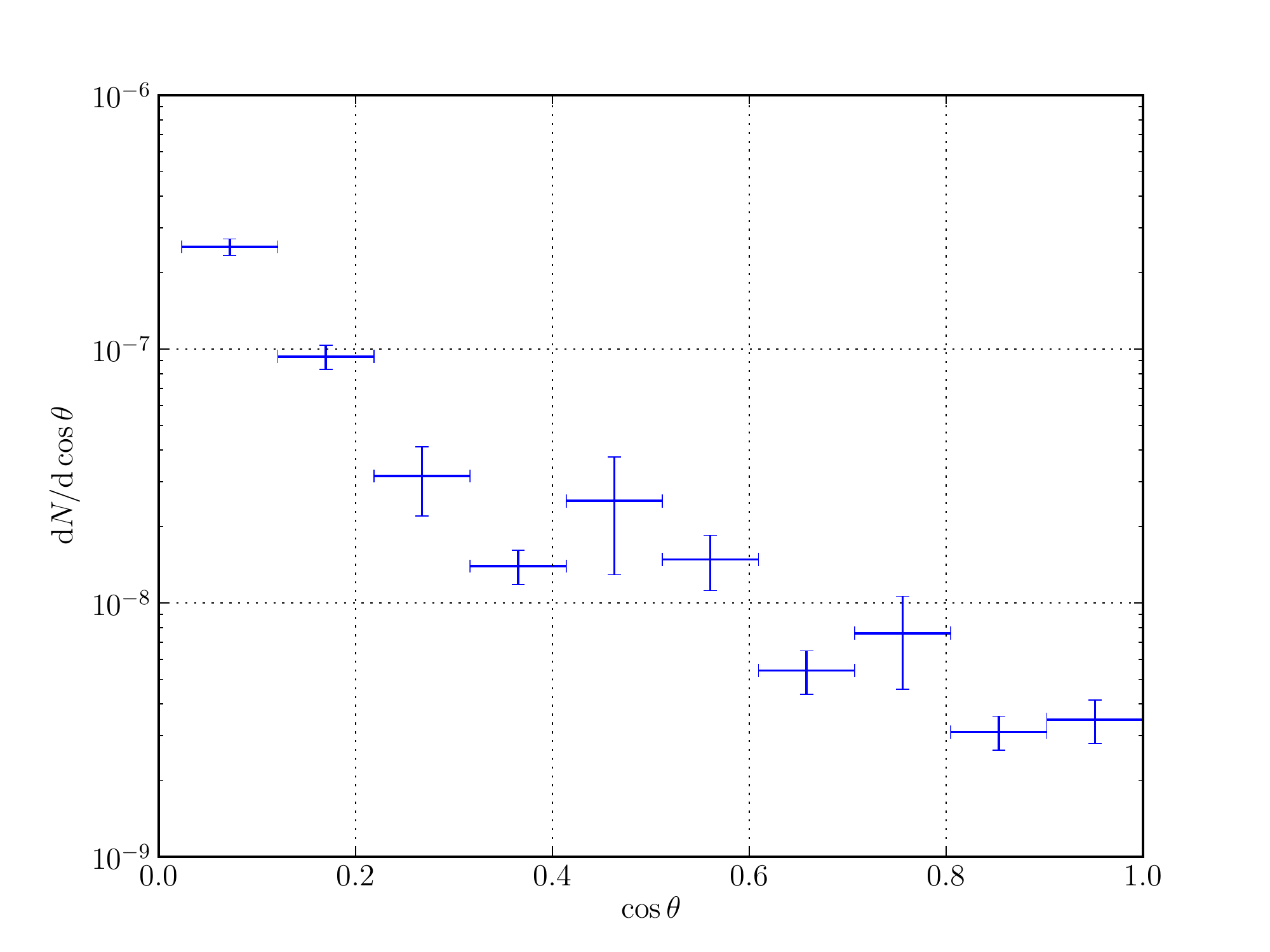}
\caption{The zenith angle distribution of pairs that would be detected in the IceCube-model detector.   Most of the pairs come from just below the horizon.}
\label{fig:zenith}
\end{center}
\end{figure}

Figure \ref{fig:zenith} shows the zenith angle distribution of accepted pairs.  As expected, most of the detected pairs are just below the horizon, where the distance between the shower and detector is smallest. For these near-horizontal neutrinos the cosmic-ray air showers have a much longer propagation distance in the atmosphere, so the interaction to detector separation never gets too small.   The dominance of the horizontal sensitivity means that the expected rates are somewhat sensitive to the detector shape; detectors with a larger horizontal frontal area should see more neutrino pairs.   

\begin{figure}[htb]
\begin{center}
\includegraphics[width=0.5\textwidth]{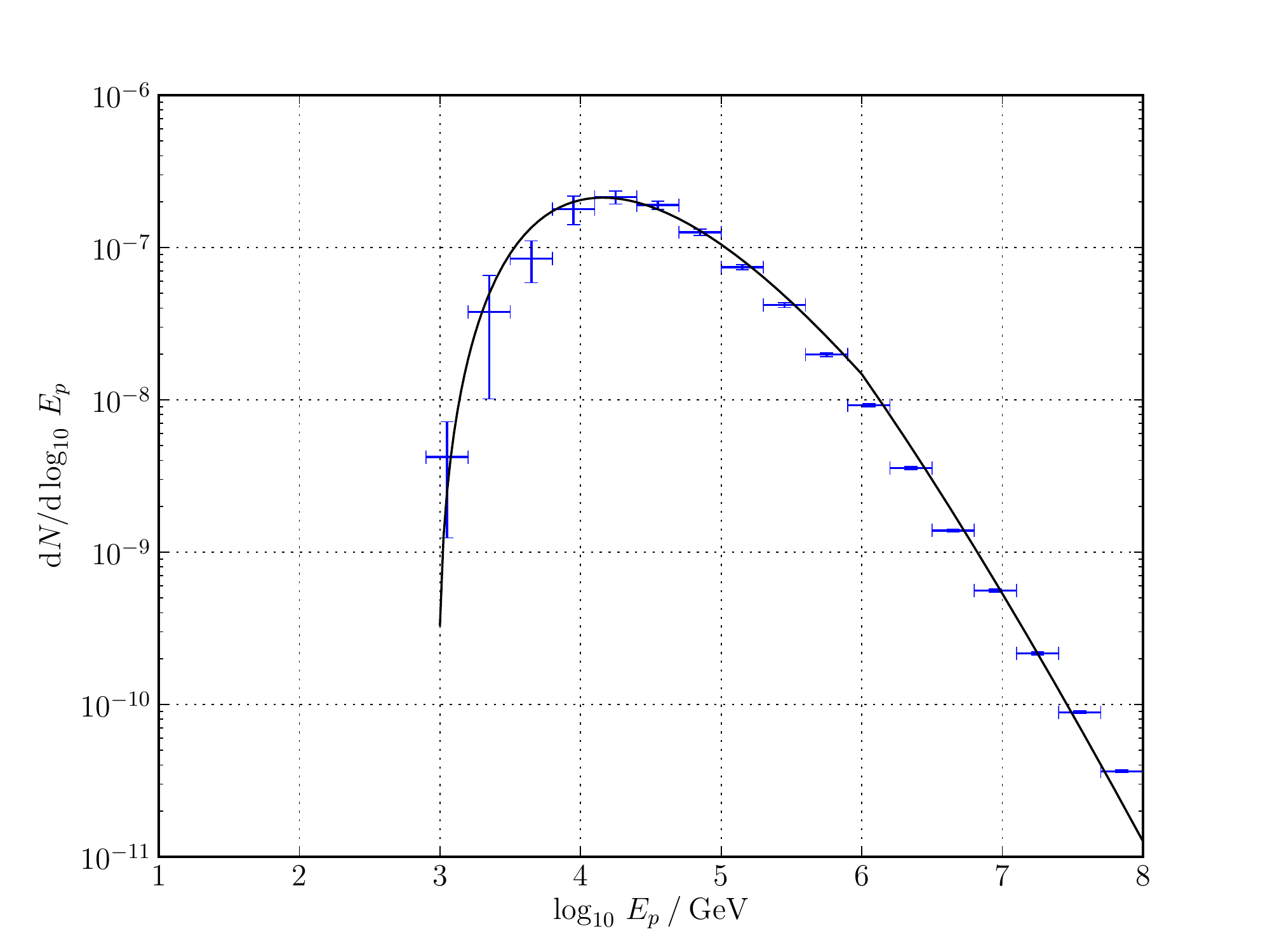}
\caption{The primary energy of the cosmic-ray progenitor of the pairs that would be detected in the IceCube-model detector. The solid black line shows the result of a fit to the %raw 
detection probability distribution, weighted with the flux $\Phi(E_p)$.}
\label{fig:primaryenergy}
\end{center}
\end{figure}

Figure \ref{fig:primaryenergy} shows the primary energies of the cosmic-ray progenitors of the pairs that would be detected.  The distribution is peaked for primaries around 30 TeV, well below the knee of the cosmic-ray spectrum, where the cosmic-ray composition is mostly protons. This peak reflects several factors: the cosmic-ray flux decrease with increasing energy, the increasing neutrino production and detection cross-sections, and the narrowing of the average opening angle with increasing neutrino energy.    With the rapid fall-off with increasing energy, uncertainties in the cosmic-ray flux at high energies, above the knee, are unimportant.   

\begin{figure}[htb]
\begin{center}
\includegraphics[width=0.5\textwidth]{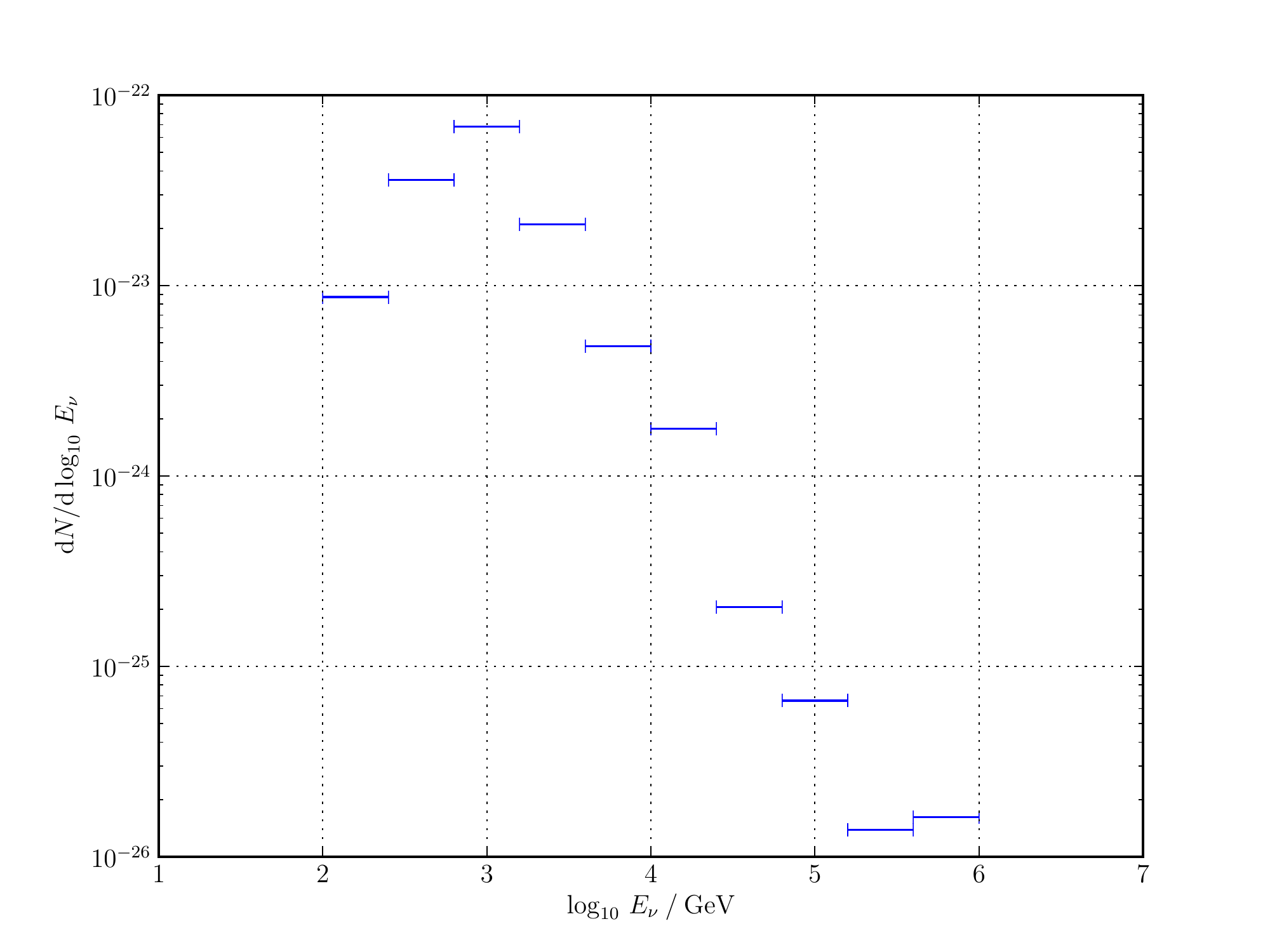}
\caption{The neutrino energies of the pairs that would be detected in the IceCube-model detector (2 entries/pair).
\label{fig:neutrinoenergy}
}
\end{center}
\end{figure}

Figure \ref{fig:neutrinoenergy} shows the energy of the observed neutrinos (with 2 entries/pair), with a 1 km maximum separation.  This distribution is peaked around 1 TeV, about 3\% of the peak of the primary energy distribution.  The maximum reflects the competition between the rapid decrease in atmospheric neutrino flux with increasing energy, the increasing interaction probability and the decreasing opening angle ($p_T/E_\nu$) with increasing neutrino energy.    Events near the minimum energy cutoff, 100 GeV, do not significantly contribute to the rate.  In this energy range, prompt neutrinos are not significant. 

The correlation between the energies of the two neutrinos is small.  This is expected, since the separation distance is determined largely by the $p_T$ and energy of the lowest energy neutrino; as long as one neutrino has an energy substantially above the other one, its energy is largely irrelevant.  

Figure \ref{fig:ratevssize} shows the predicted detection rate as a function of detector diameter.  For small detectors, the naive rate should scale as roughly the square of the surface area of the detector, or as the effective volume to the $4/3$ power.     For larger detectors, the rate increase is slower, because of the drop in neutrino flux at large transverse momentum.   

\section{Signal and Background Rates}

The overall neutrino rates for a 1 km$^3$ detector are shown in Table \ref{tab:Rates}, for both QGSJET and DPMJET, for both an all-proton and all-iron assumed cosmic-ray composition.  The rates are all in quite good agrement, with the composition making at most a 21\% differerence.  At the relevant energies, a few hundred TeV, cosmic-rays are expected to be mostly protons and lighter nuclei.

\begin{figure}[htb]
\begin{center}
\includegraphics[width=0.5\textwidth]{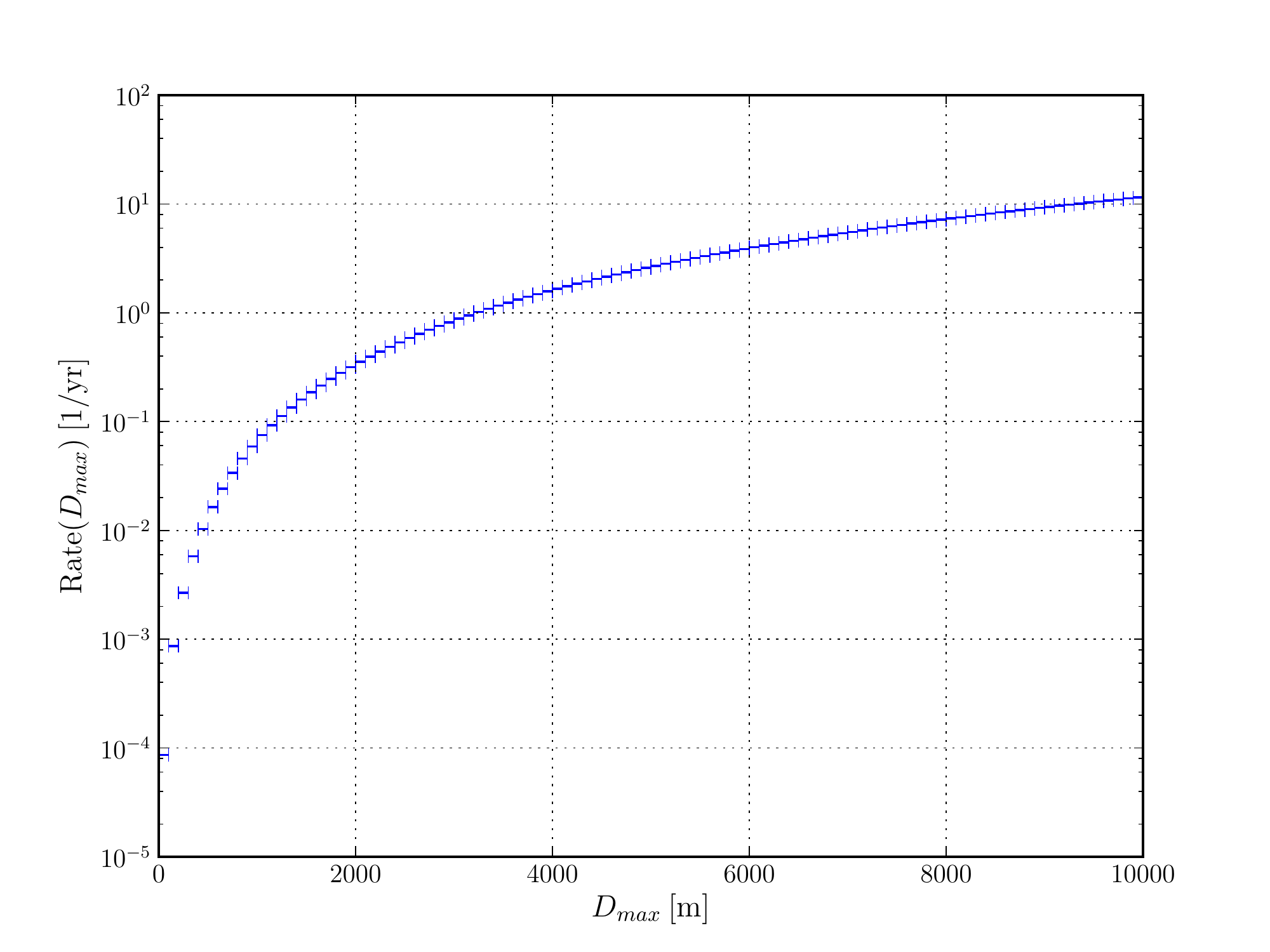}
\caption{The predicted pair-detection rate as a function of detector diameter, $D_{max}$. This is for an assumed roughly spherical detector.  IceCube fits this model fairly well.  KM3NeT is likely to be wider than it is high, so the KM3NeT rates determined here may be slight overestimates.}
\label{fig:ratevssize}
\end{center}
\end{figure}

For KM3NeT, using a 6 km$^3$ effective volume \cite{Coniglione}, the rate is about 11 times higher, or about 0.8 events/year.  KM3NeT is likely to be wider than it is high, so Fig. \ref{fig:ratevssize} may slightly overestimate its rate.
These rates do not capture the details of the either detector construction, but the IceCube rate should be accurate within 50\%.  More detailed calculations would require a complete simulation and an analysis chain.

\begin{table}\begin{center}
	\begin{tabular}{ c | r | r | }
		\cline{2-3}
								& 	\textsc{qgsjet}  [1/yr]	& 	\textsc{dpmjet}  [1/yr]		\\
		\hline
		\multicolumn{1}{| l |}{Protons}	&	$0.068$			&	$0.070$				\\
		\multicolumn{1}{| l |}{Iron}		&	$0.065$			&	$0.056$				\\
		\hline
	\end{tabular}
	\caption{The calculated event rates in IceCube for two interaction models and assumed all-proton or all-iron cosmic-ray composition.    The rates do not vary very much for the four choices.  Most of the double neutrino events have cosmic-ray progentors with energies of a few hundred TeV, where the composition is expected to be mostly protons and lighter nuclei.}
	\label{tab:Rates}
\end{center}\end{table}

The flux of double neutrino events is large enough that a signal should be visible in the proposed KM3NeT detector, and an event might be seen in IceCube.   Once events are seen, then it is necessary to try to classify them as double neutrinos or as due to new physics.   The observed specific energy loss ($dE/dx$) \cite{Abbasi:2012wht} and zenith angle distributions may help in separating the two classes of events.  As Fig. \ref{fig:neutrinoenergy} shows, about 20\% of the mons have energies above ~ 2 TeV, with an average $dE/dx$ more than 10 times minimum ionizing; this is a larger energy loss than is expected from the considerably heavier supersymmetric or Kaluza-Klein particles.  More importantly, most of the neutrino pairs come from near the horizon, whereas neutrino induced supersymmetry interactions are more evenly spread over the upward-going hemisphere.    However, the angular distribution is similar to what one would expect from supersymmetric or Kaluza-Klein particles that are produced directly in cosmic-ray air showers.  Although these identifying criteria may be inadequate to classify a single event, a small event sample should allow clear conclusions to be drawn.  

One potential background to these events (and to searches for supersymmetric and Kaluza-Klein particles) is from muon pairs that are produced in neutrino interactions (or in cosmic-ray air showers), from decays of charmed particles or Drell-Yan pairs.  This background has been discussed previously \cite{Albuquerque:2008zs}.  However, the constraint that the two tracks appear parallel is a powerful constraint to eliminate background.  For long muon tracks, IceCube has an angular resolution that is better than 1$^0$ \cite{Abbasi:2010rd}; KM3NeT is expected to be a few times better, but we will use a maximum opening angle $\theta_o = 1^0$.  Tracks pairs that diverge by more than twice that, or $2^0$ can be eliminated.  The $2^0$ parallelism requirement requires that the muons must originate at a vertex at least 2865 m from the detector; if we require that the tracks traverse through most of IceCube, this gives a minimum track length of 3800 m.  Muons travelling 3800 m  in ice must have a minimum initial muon energy of at least 6 TeV.  If the muons path is mostly rock (as in below IceCube or KM3NeT), then the energy threshold would be three times higher.   A 6 TeV muon with a $1^0$ opening angle requires $p_T = E_\mu\cdot \sin(\theta_o) = 105$ GeV/c to satisfy the track separation requirement.  Such a large $p_T$ is extremely rare; for comparison the IceCube studies of down-going muons covered the range of a few GeV/c.     The large $p_T$ is required for a range of conditions.  For vertices farther from the detector, the opening angle is smaller, but the muon energy rises more quickly, increasing the minimum $p_T$.   The inclusion of multiple scattering will alter these numbers slightly, but should not change the overall conclusion that the background rate due to neutrino interactions is negligible.  

Angular misreconstruction does not affect the conclusions very much.  As the allowed actual opening angle rises, the vertex can be closer to the detector and the minimum muon energy drops.  However, as the opening angle rises, and $p_T$  stays large.  For example, for  $\theta_o=5^0$, the vertex can be 1100 m from the detector, but the required $p_T$ is still 87 GeV.   If the distance between the two tracks were mis-reconstructed, with, effectively a smaller two-track separation requirement, significant background could be found.

Similar arguments apply for dimuons coming from cosmic-ray air showers.  For the muons to be upward-going, the air showers  muons must traverse more than 100 km of ice or water (at a depth of 1500 m, the horizontal distance to the surface is 138 km).  This is not possible, but there may still be a small background from downward-going dimuons where both muons are misreconstructed as upward-going. 

A third background is from neutrino pairs where the two neutrinos are produced by different cosmic-ray interactions.  The rate for this background depends on the detector angular and temporal resolution; it can be estiated with Poisson statistics.  IceCube observes about 50,000 muons from high energy neutrino interactions per year, spread over $2\pi$ steradians and $3\times10^7$ s. For a time difference resolution of 450 ns \cite{IceCube:2012ij}, the number of temporal overlaps is 0.0008 per year.  Including the $2^0$ paralelleism requirement reduces the rate by another factor of 6,000.  This calculation ignores non-uniformities in the acceptance, but these are not large effects.   KM3NetT is larger, but its better angular resolution should lead to a similar rate. 

\section{Conclusions}

In conclusion, the expected rate for a 1 km$^3$ detector like IceCube to observe two upward-going neutrinos from the same cosmic-ray air shower is about one every 14 years.  Future, larger detectors, like a 6 km$^3$ KM3NeT will have a substantially larger rate, i.e. 0.8 per year, and so should observe a signal.  These double-neutrino events are an irreducible background to searches for pairs of upward-going particles produced by beyond-the-standard-model processes.   The other standard-model backgrounds to these processes appear to be very small. 

We thank Lisa Gerhardt for help with the Monte Carlo simulations and numerous useful discussions and Dave Seckel  and Klaus Helbing for their insightful comments on an early draft of this paper.   This work was supported in part by U.S. National Science Foundation under grant 0653266 and the U.S. Department of Energy under contract number DE-AC-76SF00098.


\begin{thebibliography}{99}

\def\etal{{\it et al.}}

\bibitem{Halzen:2008zz} 
  F.~Halzen and S.~R.~Klein,
  Phys.\ Today {\bf 61N5}, 29 (2008).
  %%CITATION = PHTOA,61N5,29;%%

\bibitem{Gaisser:1994yf} 
  T.~K.~Gaisser, F.~Halzen and T.~Stanev,
  %``Particle astrophysics with high-energy neutrinos,''
  Phys.\ Rept.\  {\bf 258}, 173 (1995)
  [Erratum-ibid.\  {\bf 271}, 355 (1996)].
  %%CITATION = HEP-PH/9410384;%%

\bibitem{IceCube} F.~Halzen and S.~R.~Klein,  Rev.\ Sci.\ Instrum.\  {\bf 81}, 081101 (2010).

\bibitem{Dzhilkibaev:2009ja} 
  Zh.-A.~Dzhilkibaev, {\it et al.} [Baikal Collaboration], 
  %``Search for a diffuse flux of high-energy neutrinos with the Baikal neutrino telescope NT200,''
  arXiv:0909.5562.

\bibitem{Collaboration:2011nsa} 
  M.~Ageron, J.~A.~Aguilar, I.~Al Samarai, A.~Albert, F.~Ameli, M.~Andre, M.~Anghinolfi and G.~Anton {\it et al.},
  %``ANTARES: the first undersea neutrino telescope,''
  Nucl.\ Instr.\ Meth.\ A {\bf 656}, 11 (2011).

\bibitem{Sapienza:2011zzb} 
  P.~Sapienza [KM3NeT Collaboration],
  %``KM3NeT: A km**3-scale neutrino telescope in the Mediterranean Sea,''
  Nucl.\ Phys.\ Proc.\ Suppl.\  {\bf 212-213}, 134 (2011).

\bibitem{Coniglione}
R.~Coniglione for the KM3NeT Collaboration, Nucl. Instrum. \& Meth. A,
http://dx.doi.org/10.1016/j.nima.2012.11.148

\bibitem{Sullivan:2012hf} 
  G.~Sullivan [IceCube Collaboration],
  %``Results from the IceCube Experiment - Neutrino 2012,''
  arXiv:1210.4195 [astro-ph.HE].
  %%CITATION = ARXIV:1210.4195;%%

\bibitem{Albuquerque:2006am} 
  I.~F.~M.~Albuquerque, G.~Burdman and Z.~Chacko,
  %``Direct detection of supersymmetric particles in neutrino telescopes,''
  Phys.\ Rev.\ D {\bf 75}, 035006 (2007).
  %%CITATION = HEP-PH/0605120;%%

%\cite{Helbing:2011wf}
\bibitem{Helbing:2011wf} 
  K.~Helbing [IceCube Collaboration],
  %``IceCube as a discovery observatory for physics beyond the standard model,''
  arXiv:1107.5227 [hep-ex].
  %%CITATION = ARXIV:1107.5227;%%

\bibitem{Albuquerque:2009vk} 
  I.~F.~M.~Albuquerque and S.~R.~Klein,
  %``Supersymmetric and Kaluza-Klein Particles Multiple Scattering in the Earth,''
  Phys.\ Rev.\ D {\bf 80}, 015015 (2009).
  %%CITATION = ARXIV:0905.3180;%%

\bibitem{Albuquerque:2008zs} 
  I.~F.~M.~Albuquerque, G.~Burdman, C.~A.~Krenke and B.~Nosratpour,
  %``Direct Detection of Kaluza-Klein Particles in Neutrino Telescopes,''
  Phys.\ Rev.\ D {\bf 78}, 015010 (2008).
  %%CITATION = ARXIV:0803.3479;%%

\bibitem{thesis}
D. van der Drift, ``Cosmic Rays: CORSIKA predictions for separated tracks," Masters thesis, 
Technische Universiteit Eindhoven, Nov., 2012.

\bibitem{IceCube:2012ij} 
  R.~Abbasi {\it et al.}  [IceCube Collaboration],
  %``Lateral Distribution of Muons in IceCube Cosmic Ray Events,''
  Phys.\ Rev.\ D {\bf 87}, 012005 (2013).
  %%CITATION = ARXIV:1208.2979;%%

\bibitem{Aartsen:2012uu} 
  M.~G.~Aartsen {\it et al.}  [IceCube Collaboration],
  %``Measurement of the Atmospheric $\nu_e$ flux in IceCube,''
  arXiv:1212.4760 [hep-ex].
  %%CITATION = ARXIV:1212.4760;%%

\bibitem{CORSIKA} D. Heck et al., CORSIKA: A Monte Carlo Code to Simulate Extensive Air Showers, Tech. Rep. FZKA 6019 (Forschungszentrum Karlsruhe
GmbH, Karlsruhe, 1998).

\bibitem{Kalmykov:1997te} 
  N.~N.~Kalmykov, S.~S.~Ostapchenko and A.~I.~Pavlov,
  %``Quark-gluon string model and EAS simulation problems at ultra-high energies,''
  Nucl.\ Phys.\ Proc.\ Suppl.\  {\bf 52B}, 17 (1997).
  %%CITATION = NUPHZ,52B,17;%%

\bibitem{Ranft:1999qe} 
  J.~Ranft,
  %``DPMJET version II.5: Sampling of hadron hadron, hadron - nucleus and nucleus-nucleus interactions at accelerator and cosmic ray energies according to the two component dual parton model: Code manual,''
  hep-ph/9911232.
  %%CITATION = HEP-PH/9911232;%%

%\cite{Hoerandel:2002yg}
\bibitem{Hoerandel:2002yg} 
  J.~R.~Hoerandel,
  %``On the knee in the energy spectrum of cosmic rays,''
  Astropart.\ Phys.\  {\bf 19}, 193 (2003)
  [astro-ph/0210453].
  %%CITATION = ASTRO-PH/0210453;%%

\bibitem{Ambrosio:1999qu} 
  M.~Ambrosio {\it et al.}  [MACRO Collaboration],
  %``High statistics measurement of the underground muon pair separation at Gran Sasso,''
  Phys.\ Rev.\ D {\bf 60}, 032001 (1999).
  %%CITATION = HEP-EX/9901027;%%

\bibitem{Bfield}National Geophysical Data Center, 2013,
http://www.ngdc.noaa.gov/geomag/WMM/calculators.shtml

\bibitem{Ribordy:2006qd} 
  M.~Ribordy,
  %``Reconstruction of Composite Events in Neutrino Telescopes,''
  Nucl.\ Instrum.\ Meth.\ A {\bf 574}, 137 (2007)
  [astro-ph/0611604].
  %%CITATION = ASTRO-PH/0611604;%%

\bibitem{Abbasi:2012wht} 
  R.~Abbasi {\it et al.}  [IceCube Collaboration],
  %``An improved method for measuring muon energy using the truncated mean of dE/dx,''
  Nucl.\ Instrum.\ Meth.\ A {\bf 703}, 190 (2013)
  [arXiv:1208.3430 [physics.data-an]].

\bibitem{Abbasi:2010rd} 
  R.~Abbasi {\it et al.}  [IceCube Collaboration],
  %``Time-Integrated Searches for Point-like Sources of Neutrinos with the 40-String IceCube Detector,''
  Astrophys.\ J.\  {\bf 732}, 18 (2011)
  [arXiv:1012.2137 [astro-ph.HE]].
  %%CITATION = ARXIV:1012.2137;%%

\end{thebibliography}
\end{document}